\pdfoutput=1
\documentclass[10pt]{iopart}
\usepackage{iopams}  
\usepackage{hyperref}
\usepackage{graphicx}

\usepackage{xcolor}
\usepackage[normalem]{ulem}

\usepackage{cite}

\usepackage{mathrsfs}
\DeclareSymbolFontAlphabet{\mathrsfs}{rsfs}
\newcommand{\scri}{\mathrsfs{I}}
\newcommand{\scrip}{$\scri^+$}

\newcommand{\tder}{\Pi}

\begin{document}

\title[Decay of solutions of the wave equation in cosmological spacetimes]{Decay of solutions of the wave equation in cosmological spacetimes - a numerical analysis}

\author{Flavio Rossetti$^{1}$ and Alex Vañó-Viñuales$^{2}$}
\address{$^1$CAMGSD, Departamento de Matemática, Instituto Superior Técnico IST, Universidade de Lisboa UL, Avenida Rovisco Pais 1, 1049-001 Lisboa, Portugal}
\address{$^2$Centro de Astrof\'{\i}sica e Gravita\c c\~ao - CENTRA, Departamento de F\'isica, Instituto Superior T\'ecnico IST, Universidade de Lisboa UL, Avenida Rovisco Pais 1, 1049-001 Lisboa, Portugal}
\ead{flavio.rossetti@tecnico.ulisboa.pt}

\begin{abstract}
We numerically evolve spherically symmetric solutions to the linear wave equation on some expanding Friedmann-Lema{\^i}tre-Robertson-Walker (FLRW) spacetimes and study the respective asymptotics for large times. We  find a quantitative relation between the expansion rate of the underlying background universe and the decay rate of linear waves, also in the context of spatially-hyperbolic spacetimes, for which rigorous proofs of decay rates are not generally known.
A prominent role in the decay mechanism is shown to be played by \textit{tails}, i.e. scattered waves propagating in the interior of the lightcone.
\end{abstract}

%
%
%
%
%

\section{Introduction}

The propagation and the decay of linear waves in cosmological backgrounds not only capture the key properties of the linearized Einstein equations, paving the way to the analysis of  non-linear stability problems, but also encompass many important features of the spacetime at hand. On the one side, the late-time asymptotics can play a role to assess the validity of the cosmic no--hair conjecture for several solutions of Einstein's equations \cite{AndreassonRingstrom, CostaNatarioOliveira}, thus giving insights on their causal structure. On the other hand, non--zero curvature causes wave scattering and is therefore an obstruction to the \textit{Huygens principle}, i.e. the propagation of information on the surface of expanding spheres. The latter phenomenon, typical of $(3+1)-$dimensional Minkowski spacetime, is indeed non-generic \cite{Guenther}. Heuristically, when the Huygens principle fails to hold, any observer living on a timelike worldline who receives a signal from a localized source will keep perceiving a fading component of such signal for all future values of the observer's proper time. It is then said that \textit{wave diffusion} occurs or, in other words, information spreads throughout the entire volume of expanding balls.  In such case, backscattered waves (\textit{tails}) forbid sharp propagation of signals by spreading in the interior of the lightcone  \cite{FaraoniSonego}.

In this work, we are concerned with the analysis of test scalar fields solving
\begin{equation} \label{wave_eqn}
\square_g \phi = 0,
\end{equation}
with respect to the underlying $(3+1)$--dimensional spacetime metric $g$ describing expanding cosmological (FLRW) universes of constant spatial curvature $K=-1$ (hyperbolic case) and $K=0$ (flat case). These are the simplest solutions of the Einstein equations that can account for the homogeneity and isotropy that we experience in our universe at large scales.   We will restrict our analysis to the case $\Lambda = 0$, due to the existence of closed-form expressions for the solutions to the Friedmann equations. Recent developments on the problem can be found in \cite{AbbasiCraig, CostaNatarioOliveira, NatarioRossetti, Ringstrom, YagdjianGalstian, DafermosRodnianski, StarkoCraig, Schlue, KlainermanSarnak, NatarioSasane, Speck, Mondal2, Mondal3} and references therein, both for the case $\Lambda = 0$ and for a non-zero cosmological constant. It has been settled, in particular,
that the decay of the energy associated to waves, caused by the redshift effect, is only one of the many factors that influence the propagation, together for instance with the non-compactness of the space sections and the dispersive properties of the background spacetime.  While we are interested in the asymptotics of solutions for large times, several approaches to study their behaviour near the Big Bang singularity have also been developed in the last years \cite{GiraoNatarioSilva, AlhoFournodavlosFranzen, FajmanUrban, Ringstrom2}. 

We will often refer to the companion paper \cite{NatarioRossetti}, where explicit expressions and decay rates for the solutions of \eref{wave_eqn} in a class of FLRW spacetimes are provided  by use of analytical techniques. In the present work, we provide evidence for the sharpness of the known estimates and complement the results via numerical methods. In particular, we obtain uniform--in--space decay rates in the case of a family of spacetimes of hyperbolic spatial curvature, for which explicit expressions for wave solutions are not currently known.  For both the spatially-flat and the spatially-hyperbolic classes of spacetimes under investigation, we also analyse the consequences of the lack of Huygens' principle by obtaining the decay rates of tails and showing their contribution to the global asymptotics.

Finally, in the dust-filled universe with hyperbolic spatial sections, the numerical methods that we adopted confirm the decay rate $t^{-\frac32}$ showed in \cite{NatarioRossetti}, and thus provide a numerical counterexample to the rate of $t^{-2}$ stated in \cite{AbbasiCraig}.

 The paper is structured as follows. In \sref{theory} we explicitly introduce the spacetimes that we consider in our experiments and derive the corresponding equations of motion for the scalar field, whose implementation is described in \sref{impl}. Section \ref{resul} shows and explains the results obtained both for the flat and hyperbolic cases. Finally, we close with some concluding remarks.

\section{Wave equation on FLRW backgrounds}\label{theory}

We analyse the initial value problem for the wave equation on the FLRW spacetimes
\begin{equation} \label{general_metric}
g = -dt^2 + a(t)^2 d\Sigma_3^{\, 2} = a^2(\tau) \left(-d\tau^2 + d\Sigma_3^{\, 2} \right), 
\end{equation}  
for some \emph{scale factor} $a(t)$ and a three-dimensional Riemannian metric $d\Sigma_3^{\, 2}$. We define the \emph{conformal time}, up to a constant, as 
\begin{equation*}
\tau = \int\frac{dt}{a(t)}.
\end{equation*}
This work is divided into two main sections.
First, we consider {\bf spatially--flat} Friedmann metrics \eref{general_metric} with scale factor $a(t)=t^p$, where $p \ge 0$, and where $d\Sigma_3^{\, 2}$ is the standard metric on $\mathbb{R}^3$. The exponent $p$ can be related to the square of the speed of sound $w$ appearing in the linear equation of state $p_m = w \rho_m$, by 
\begin{equation*}
p = \frac{2}{3(1+w)}.
\end{equation*}
 Here, $p_m$ and $\rho_m$ are the pressure and the energy density, respectively, of the perfect fluid which constitutes the matter content of the FLRW universe (see also appendix A in \cite{NatarioRossetti}).
In this context, we evolve solutions starting from smooth, spherically symmetric initial data, and recover the results showed in \cite{NatarioRossetti}, while explicitly describing the dynamics inside and outside the lightcone centred at the origin of the coordinate system. These results are shown in subsection \ref{resulflat}.

Secondly, we repeat the above analysis for {\bf spatially--hyperbolic} metrics \eref{general_metric} endowed with scale factor
\begin{equation} \label{scalefactor_hyp}
a(\tau) = \left[\sinh \left(\frac{3w+1}{2}\tau \right) \right]^{\frac{2}{3w+1}},
\end{equation}
where $w \ge -1$, $w \ne -\frac13$, is again the square of the speed of sound of the  FLRW perfect fluid.  Here $d\Sigma_3^{\, 2}$ corresponds to the standard line element \begin{equation}\label{hyperb_spatial_line_element}
dr^2 + \sinh(r)^2\left ( d\theta^2 + \sin^2(\theta) d\varphi^2 \right)
\end{equation}
on the hyperbolic space $\mathbb{H}^3$.
 Differently from the flat case, all scale factors under consideration here reveal the same exponentially-increasing asymptotic behaviour (when represented using the conformal time coordinate). In this context, we exhibit a new pattern of decay rates in function of the parameter $w$, and recover the behaviours described in \cite{NatarioRossetti} for the dust ($w=0$) and radiation ($w=\frac13$) cases, for which an explicit formula for the solution could be obtained. We show these results in subsection \ref{resulhyperb}.

To perform the analysis, we use the physical time $t$ and the definition of the Laplace-Beltrami operator $\square_g$ to write
\begin{equation} \label{laplace_beltrami}
\square_g \phi  = \frac{1}{\sqrt{-g}}\partial_{\mu} \left(\sqrt{-g} g^{\mu \nu} \partial_{\nu} \phi \right) = - \partial_t^2 \phi - 3 \frac{\partial_t a}{a} \partial_t \phi + \frac{1}{a^2} \Delta_{\Sigma_3} \phi,
\end{equation}
with $\Delta_{\Sigma_3}$ being the Laplace-Beltrami operator defined on $\Sigma_3 = \mathbb{R}^3, \mathbb{H}^3$. When $a(t) \equiv 1$, the problem reduces to studying the linear wave equation on Minkowski or on a ultra-static spacetime with hyperbolic spatial curvature, respectively. 

From now on, we will consider \textbf{spherically symmetric} solutions $\phi = \phi(t, r)$, to facilitate the numerical study. Their behaviour is not expected to depart significantly from that of general solutions, due to the homogeneity and isotropy of FLRW spacetimes.

\subsection{Flat case}
For $K=0$, the wave equation \eref{wave_eqn} becomes
\begin{equation} \label{wave_flat_physical}
-\partial_t^2 \phi -3 \frac{\partial_t a}{a}\partial_t \phi + \frac{1}{a^2}\left(\partial_r^2 \phi + \frac2r \partial_r \phi \right) = 0
\end{equation}
after using \eref{laplace_beltrami}, where $a=a(t)$ and $\phi=\phi(t, r)$ are functions of the physical time $t$. We stress that $u=\tau-r$ and $v = \tau + r$ are null coordinates for \eref{general_metric}. So,
in order to study the decay mechanism inside and outside the lightcone $\{ |x| < \tau \}$, it will be useful to consider the wave equation written in terms of the conformal time coordinate $\tau$:
\begin{equation} \label{wave_flat_conformal}
-\partial_{\tau}^2 \phi -2 \frac{\partial_{\tau} a}{a}\partial_{\tau} \phi +\left(\partial_r^2 \phi + \frac2r \partial_r \phi \right) = 0,
\end{equation}
now having that $a=a(\tau)$ and $\phi=\phi(\tau, r)$. 

\subsection{Hyperbolic case} 
For $K=-1$, we can again use the conformal time coordinate to evolve the wave equation \eref{wave_eqn} as
\begin{equation} \label{wave_hyp_conformal}
-\partial_{\tau}^2 \phi -2 \frac{\partial_{\tau} a}{a}\partial_{\tau} \phi +\left(\partial_r^2 \phi + 2 \coth(r) \partial_r \phi \right) = 0.
\end{equation}
Since $\tau \sim \log t$ for every FLRW solution under inspection, evolving in conformal time permits us to analyse the behaviour of waves for exponentially large time intervals, after converting the results back to the physical time coordinate.

\section{Implementation}\label{impl}

The wave equation \eref{wave_eqn} is expressed as a first-order-in-time and second-order-in-space system by introducing the quantity $\tder = \partial_t\phi$  or $\tder = \partial_\tau\phi$, so the actual equations implemented are
\begin{equation}
\fl\textrm{Physical t:}\left\{\begin{array}{l}\partial_t\phi = \tder , \\ \partial_t\tder = - 3 \frac{\partial_t a}{a} \tder + \frac{1}{a^2} \Delta_{\Sigma_3} \phi ,\end{array}\right. \  
\textrm{Conformal $\tau$:}\left\{\begin{array}{l}\partial_\tau\phi = \tder , \\ \partial_\tau\tder = - 2 \frac{\partial_\tau a}{a} \tder + \Delta_{\Sigma_3} \phi .\end{array}\right.
\end{equation}

Both physical and conformal time coordinates were tested, but we adopt the latter formalism for the hyperbolic case. This is to avoid expressing the scale factor \eref{scalefactor_hyp} in physical time, process that generally requires a numerical inversion of the function $t(\tau)$. All final decay rates are converted back to the physical time coordinate for the analysis and representation of the results.

The numerical experiments were performed with a spherically symmetric code that uses the Method of Lines, with 4th order finite differences for the radial derivatives and a 4th order Runge-Kutta method for the  explicit time integration. Kreiss-Oliger dissipation \cite{kreiss1973methods,Babiuc:2007vr} is added to the right-hand-sides of the equations for stability purposes.  Different values of dissipation were tested in order to  verify that this artificial term does not affect the final decay results.  

The initial data is a Gaussian pulse centered at the origin of the form
\begin{equation}\label{inidata}
\begin{array}{ll}
\phi_0 \hspace{-0.7em} &= A e^{-\frac{r^2}{2\sigma^2}},\\
\tder_0 \hspace{-0.7em} &= v\, \partial_r\phi_0 , 
\end{array}
\end{equation}
with $A, \sigma$ and $v$ constant parameters, where $v$ gives us some control over the direction that the pulse begins propagating towards. 
Although the above initial data is not compactly supported, effectively it is after a suitable choice of the parameters, due to its rapid decay and to the machine precision to be taken into account. Considering compact-support-like initial data lets us compare our results with those present in the literature.


In the post-processing of the data, the procedure we follow to visualize the power $\alpha$ of the decay rate, for $\phi\sim t^{\alpha}$ (and appropriate changes are made when $\phi \sim (\log t)^{\alpha}$), is to plot $\frac{d\log\phi}{d\log t}\sim\alpha$ as a function of $\log t$. This is used in figures~\ref{fig:flat_phi_globaldecay}, \ref{fig:flat_phi_decayp1}, \ref{fig:flat_phi_insidelightcone} and \ref{fig:hyp_phi_globaldecay}. 

\section{Results}\label{resul}

\subsection{Flat case}\label{resulflat}
\begin{figure}[h!]
\centering
\includegraphics[width=\textwidth]{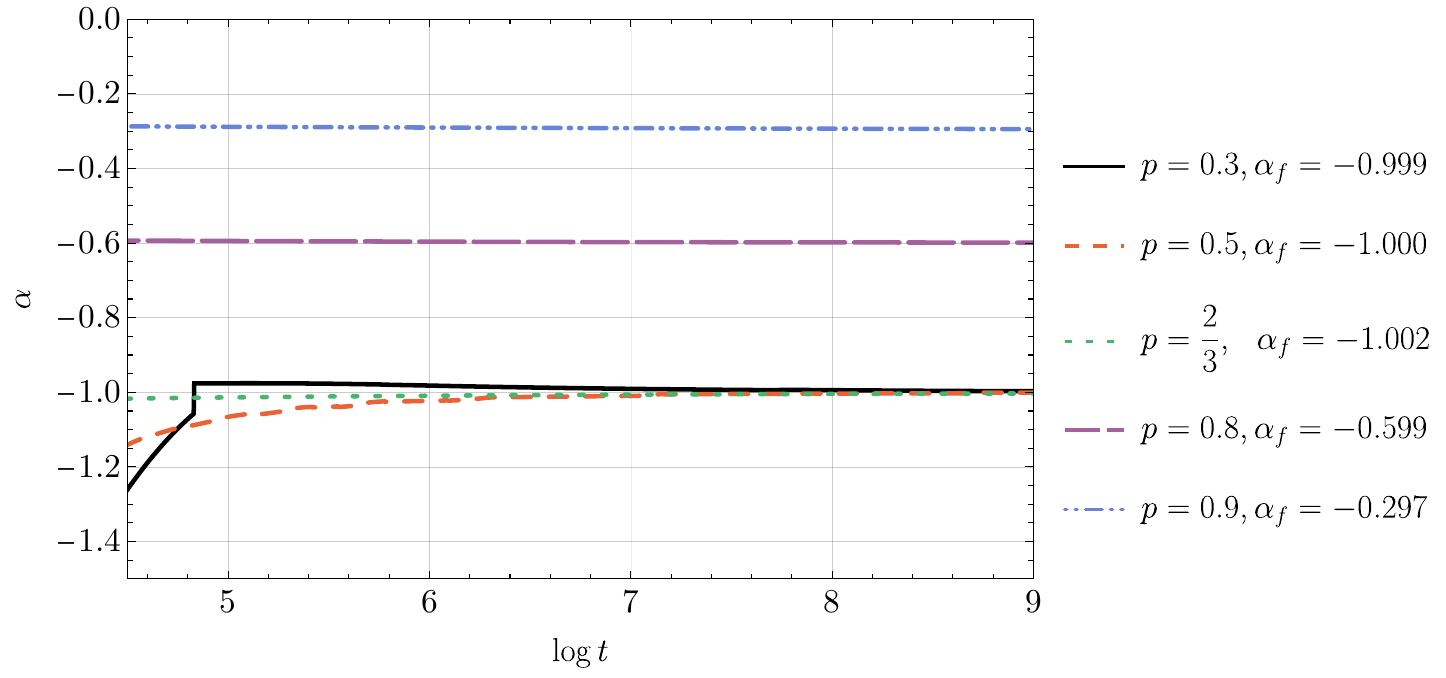}
\caption{$\sup_{\mathbb{R}^3}|\phi|$ decays as $t^{\alpha}$ for $0 < p < 1$. Initial data \eref{inidata}, $dt=0.01$. For $p=0.3, 0.5$: $dr=0.5$, $r \in [0, 2000]$. For $p=2/3, 0.8, 0.9$: $dr=0.05$, $r \in [0, 100]$. The value $\alpha_f$ represents the final value of each plotted line. The early-time behaviour is due to the interplay between redshift and dispersion and is explained in detail in section \ref{resulhyperb}, where the same phenomenon occurs in a more evident way.} 
\label{fig:flat_phi_globaldecay}
\end{figure}

\begin{figure}[h!]
\centering
\includegraphics[width=0.75\textwidth]{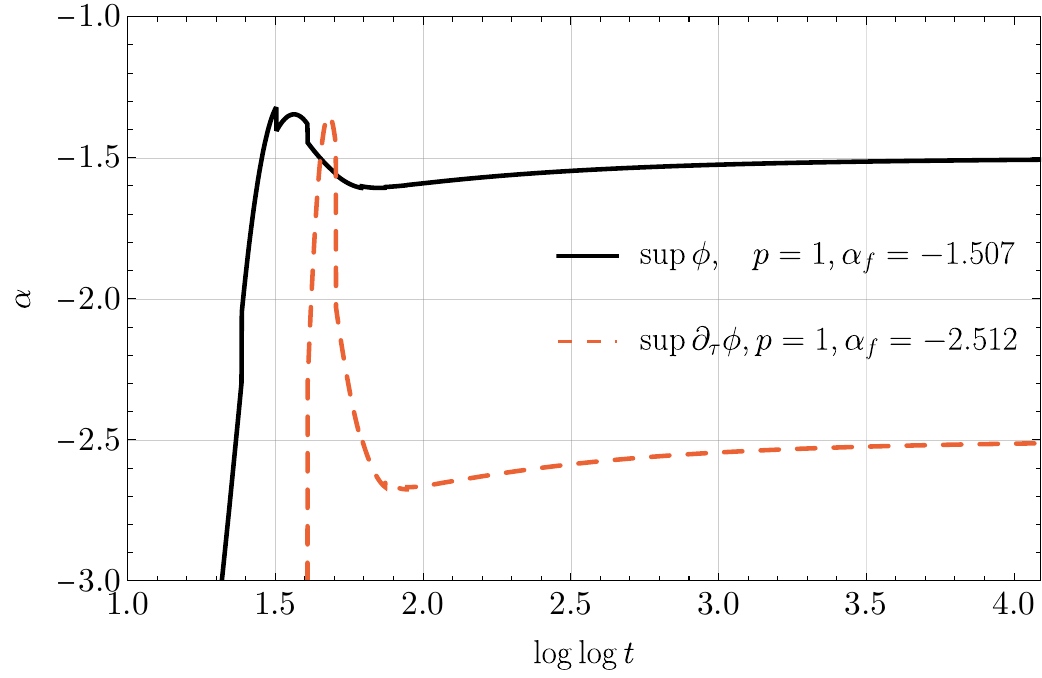}
\caption{$\sup_{\mathbb{R}^3}|\phi|$ and $\sup_{\mathbb{R}^3}|\partial_{\tau} \phi|$ decay as $(\log t)^{\alpha}$, when $p=1$. Notice that $a(t)\partial_t \phi = \partial_{\tau} \phi$. Initial data \eref{inidata}, $dt=0.01$, $dr=0.05$, $r \in [0, 100]$. The value $\alpha_f$ represents the final value of each plotted line.}
\label{fig:flat_phi_decayp1}
\end{figure}

We will now focus on solutions to \eref{wave_flat_physical} with $a(t) = t^p$, thus evolving on the background metric 
\begin{equation*}
g = -dt^2 +t^{2p} \left((dx^1)^{\, 2}  + (dx^2)^{\, 2} + (dx^3)^{\, 2} \right).
\end{equation*}
Our numerical experiments allowed us to study the maximum of the absolute value of the fields as a function of time, and returned results that agree with the following decay rates. As is shown on figures~\ref{fig:flat_phi_globaldecay} and \ref{fig:flat_phi_decayp1}, where some explanatory cases are presented, we have obtained: 
\renewcommand*{\arraystretch}{1.2}
\begin{equation} \label{flat_decay_phi}
 \sup_{x \in \mathbb{R}^3}|\phi|(t, x) \sim \left\{
\begin{array}{ll}
t^{-1}, &\textrm{if } 0 \le p \le \frac23, \\
t^{3p-3}, &\textrm{if }\frac23 \le p < 1, \\
(\log t)^{-\frac32}, &\textrm{if }p = 1,
\end{array}
\right.
\end{equation}
and:
\begin{equation} \label{flat_decay_pi}
 \sup_{x \in \mathbb{R}^3}|\partial_t \phi|(t, x) \sim \left\{
\begin{array}{ll}
t^{-1-p}, &\textrm{if }0 \le p \le \frac34, \\
t^{3p-4}, &\textrm{if }\frac34 \le p < 1, \\
t^{-1}(\log t)^{-\frac52}, &\textrm{if }p = 1,\\
t^{1-2p}, &\textrm{if }p > 1.
\end{array}
\right.
\end{equation}
The solution $\phi$ does not decay to zero for $p > 1$, however, by integrating the respective decay rate in \eref{flat_decay_pi}, $\phi$ can be seen to converge to the initial data $\phi_0$ (see \cite{CostaNatarioOliveira}).

 We emphasize that, if $\phi$ solves the wave equation $\square_g \phi = 0$, then $\partial_t \phi$ does not generally solve the same equation. It is not then surprising that the intervals of $p$ determining the decays in \eref{flat_decay_pi} differ from the intervals in \eref{flat_decay_phi}.

Among the above results, the Minkowski spacetime ($p=0$) exhibits the influence of dispersive effects on waves: here, for large times, linear waves decay to the zero solution due to their dispersion in a non-compact space \cite{Sogge}. Furthermore, if $\phi$ solves the wave equation in Minkowski spacetime, so does $\partial_t \phi$, which therefore decays with the same rate.  In expanding spacetimes, dispersive effects are weaker, since spatial regions are moving farther apart, and redshift comes into play.  A further analysis of the correlation between the decay of waves and the expansion rate of the underlying background in the flat case, also based on physical grounds, can be found in the introduction of \cite{NatarioRossetti}. We stress that such a correlation does not hold in the hyperbolic case studied in section \ref{resulhyperb}: although all spacetimes in question have scale factors characterized by the same asymptotic behaviour for large times, the respective decay rates are generally different. The behaviour of tails, on the other hand, does not differ between the flat and the hyperbolic case. This suggests that the behaviour of backscattered waves encodes the relation between redshift and dispersion at a more fundamental level. We now analyse such phenomenon. 

\begin{figure}[h!]
\centering
\includegraphics[width=\textwidth]{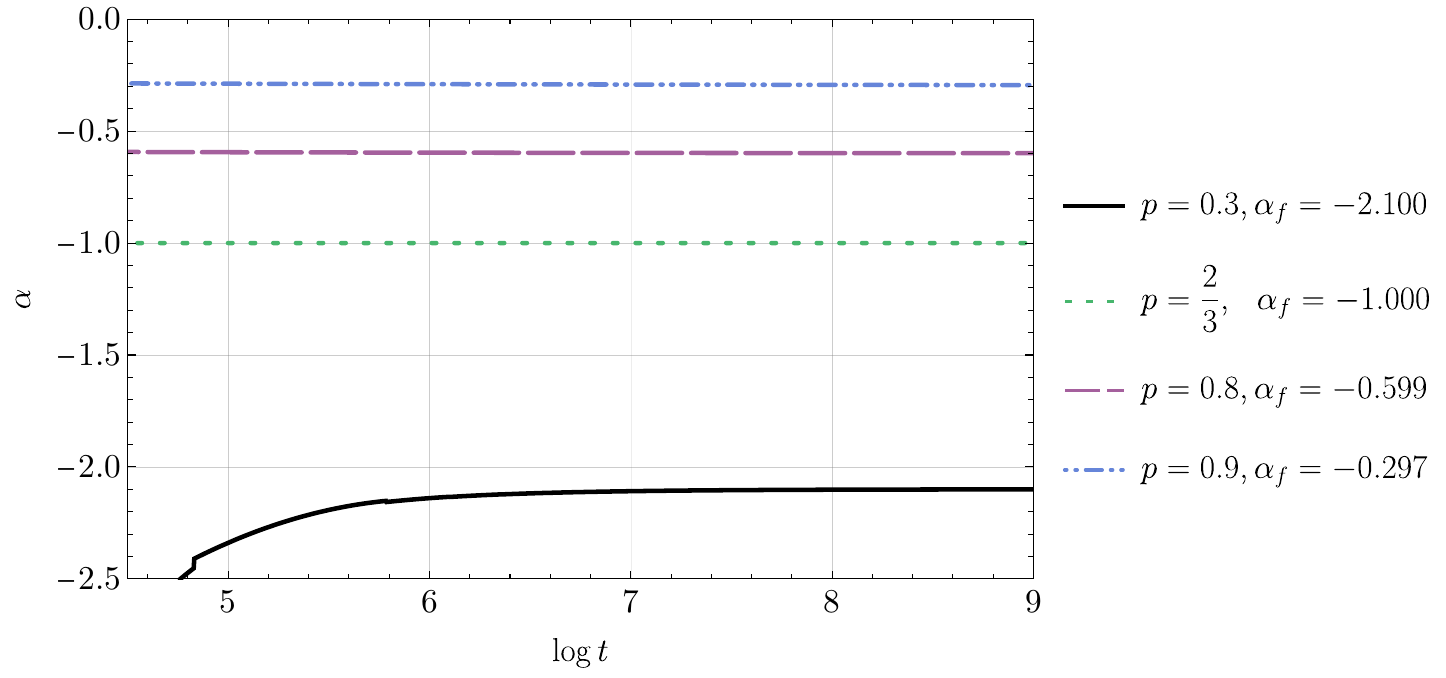}
\caption{Inside the lightcone, flat case: $\sup_{\{2r < \tau\}} |\phi|$ decays as $t^{\alpha}$. Initial data \eref{inidata}, $dt=0.01$. For $p=0.3$: $dr=0.5$, $r \in [0, 2000]$. For $p=2/3, 0.8, 0.9$: $dr=0.05$, $r \in [0, 100]$. The value $\alpha_f$ represents the final value of each plotted line.}
\label{fig:flat_phi_insidelightcone}
\end{figure}

\begin{figure}[h!]
\centering
\begin{minipage}{.5\textwidth}
  \centering
  \includegraphics[width=0.95\linewidth]{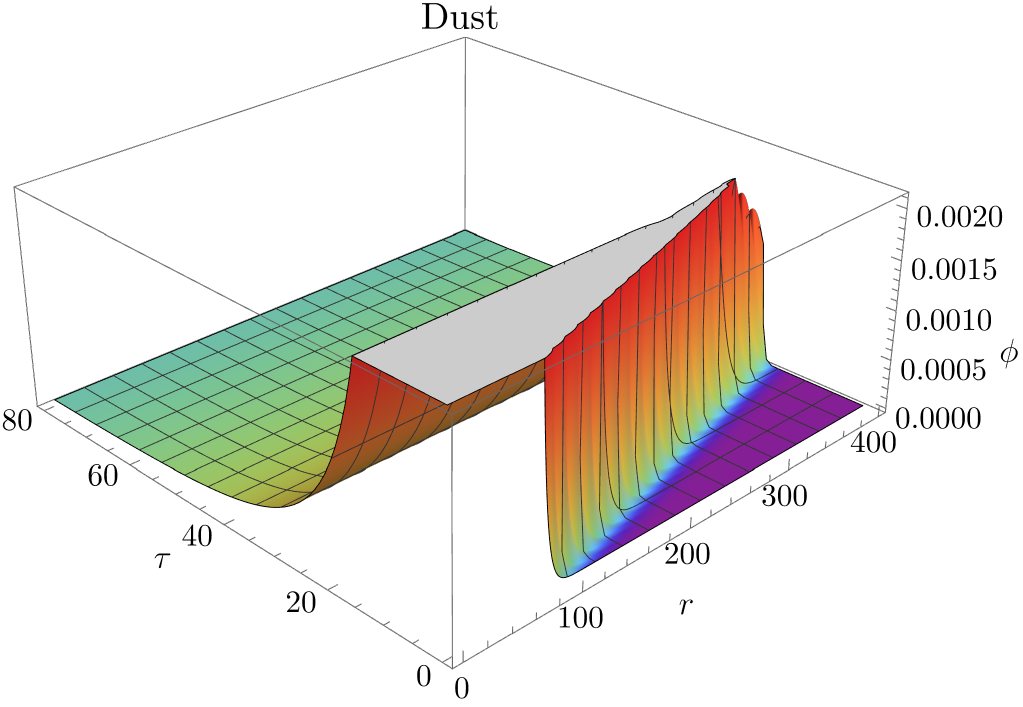} 
  
\end{minipage}%
\begin{minipage}{.5\textwidth}
  \centering
  \includegraphics[width=0.95\linewidth]{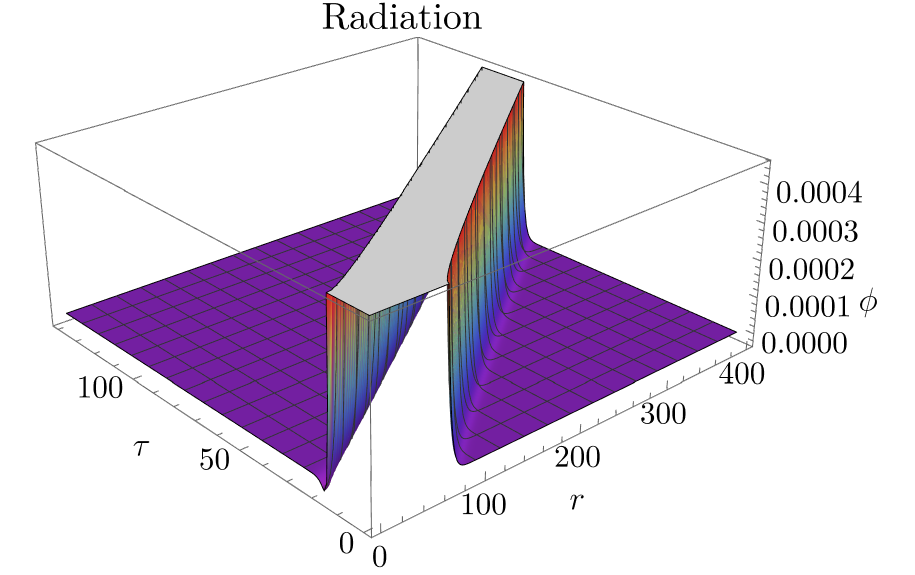} 
  
\end{minipage}
\caption{Plots of $\phi$ as a function of time and radius for the flat case. Different colors correspond to different values of the function. For the dust-filled universe ($p=\frac23$), a non-zero contribution given by tails can be observed inside the lightcone. In the radiation case ($p=\frac12$), the Huygens principle holds. We truncated the range of the values of $\phi$ for clarity.} \label{fig:3Dplots}
\end{figure}

For any $p \in (0, 1), p \ne \frac12$, we have, for every $\delta \in (0, 1)$ (see also figure \ref{fig:flat_phi_insidelightcone}):
\begin{equation*}
\sup_{ |x|<\delta \tau } |\phi| \sim 
 t^{3p-3}
\end{equation*}
and
\begin{equation*}
 \sup_{|x|<\delta \tau } |\partial_t \phi| \sim
t^{3p-4}.
\end{equation*}
These represent the decay of the backscattered tails inside the lightcone, and their non-zero contributions reveal the failure of Huygens' principle. We stress that for $p=0$ (Minkowski) and $p=\frac12$ (radiation) the Huygens principle holds in $3+1$ dimensions. Therefore, the restriction of $\phi(\tau, \cdot)$ to the lightcone $\{|x| < \delta \tau \}$ will eventually be zero for large times (compare with figure \ref{fig:3Dplots}).  In the radiation-filled universe, the validity of the principle follows from the vanishing of the four-dimensional Ricci scalar. In fact, the wave equation in such a spacetime is equivalent to the conformally invariant wave equation, which, in turn, implies that $\square_{\eta} (\tau \phi)= 0$ when $\eta$ is the metric of Minkowski's spacetime \cite{NatarioRossetti}. The Huygens principle is expected to be an unstable phenomenon. In particular, in vacuum, it holds only in Minkowski space and in plane-wave spacetimes \cite[Chapter VIII, Theorem 1.4]{Guenther}.  An integral equation describing the solution of the wave equation in terms of the values assumed on a past lightcone was found in \cite{Mondal1} and can be applied to spacetimes for which the Huygens principle does not hold.

\begin{figure}[h!]
\centering
\includegraphics[width=0.75\textwidth]{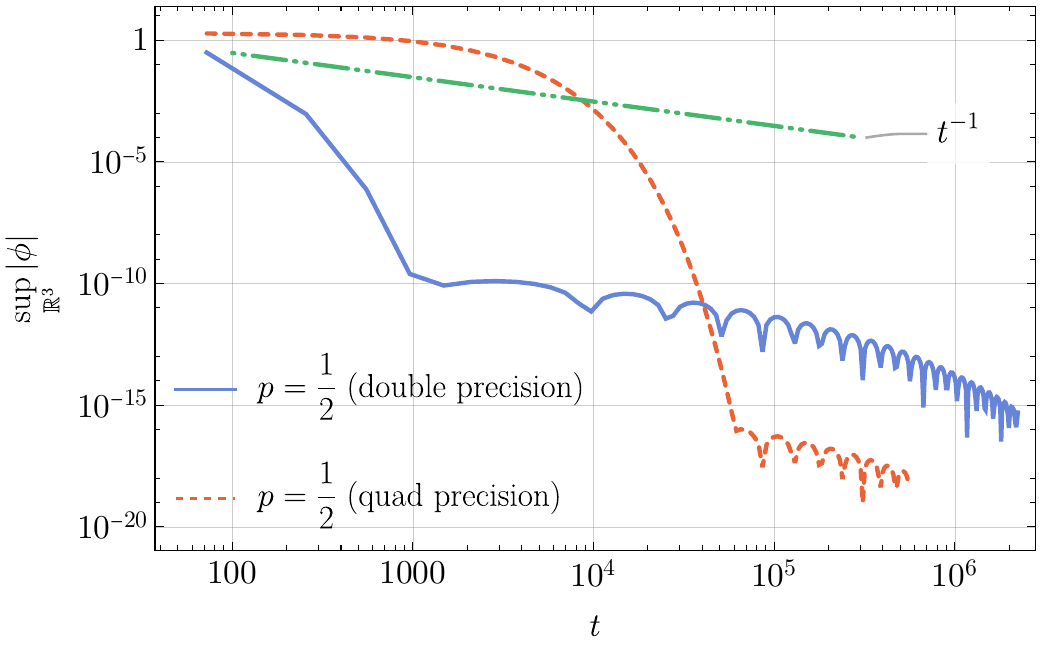}
\caption{Inside the lightcone, flat case: behaviour of $\sup_{\{6r < \tau \}} |\phi|$ for $p=0.5$. The solution is eventually identical to zero in this region due to Huygens' principle, but some non-zero contributions will be detected depending on the machine precision. Initial data \eref{inidata}, $dt=0.01$, $dr=0.5$, $r \in [0, 3000]$.}
\label{fig:flat_phi_plotp05huygens}
\end{figure}

Isolating the decay rate of tails from the global asymptotics of wave solutions requires some care. In order to observe the decay of tails, it is necessary to evolve simulations for a long enough time so that, by then, the outgoing waves will have left the interior of the lightcone.  At the same time, this task requires to monitor the smallness of the solution $\phi$. The non-zero contributions, indeed, should not be confused with those caused by finite machine precision (see figure \ref{fig:flat_phi_plotp05huygens} for a visualization of this effect). This problem is even more challenging in the hyperbolic case, where solutions expressed in conformal time have an exponentially fast decay. 

Outside such a lightcone, the asymptotics for $\phi$ are the same as in \eref{flat_decay_phi} and the ones for $\partial_t \phi$ are as in \eref{flat_decay_pi}.
In particular, the decay occurring in a region strictly contained inside the lightcone becomes dominant for $\frac23 \le p < 1$. In fact, a rapid expansion of the spatial sections (the larger the $p$, the faster the expansion) increases the effect of redshift on the waves that propagate along the outgoing direction. On the other hand, dispersive effects decrease and thus the backscattered waves gather most of the total energy density, therefore providing the main contribution to the decay. For $p > 1$ the causal structure of the FLRW solution changes (future null infinity \scrip$ $ becomes spacelike) and so does the behaviour of waves. We also stress that the pointwise decay of solutions does not generally reveal the presence of tails: such phenomenon is visible when studying the $L^{\infty}$--norm decay of solutions.

\subsection{Hyperbolic case} \label{resulhyperb}
\begin{figure}[h!]
\centering
\includegraphics[width=\textwidth]{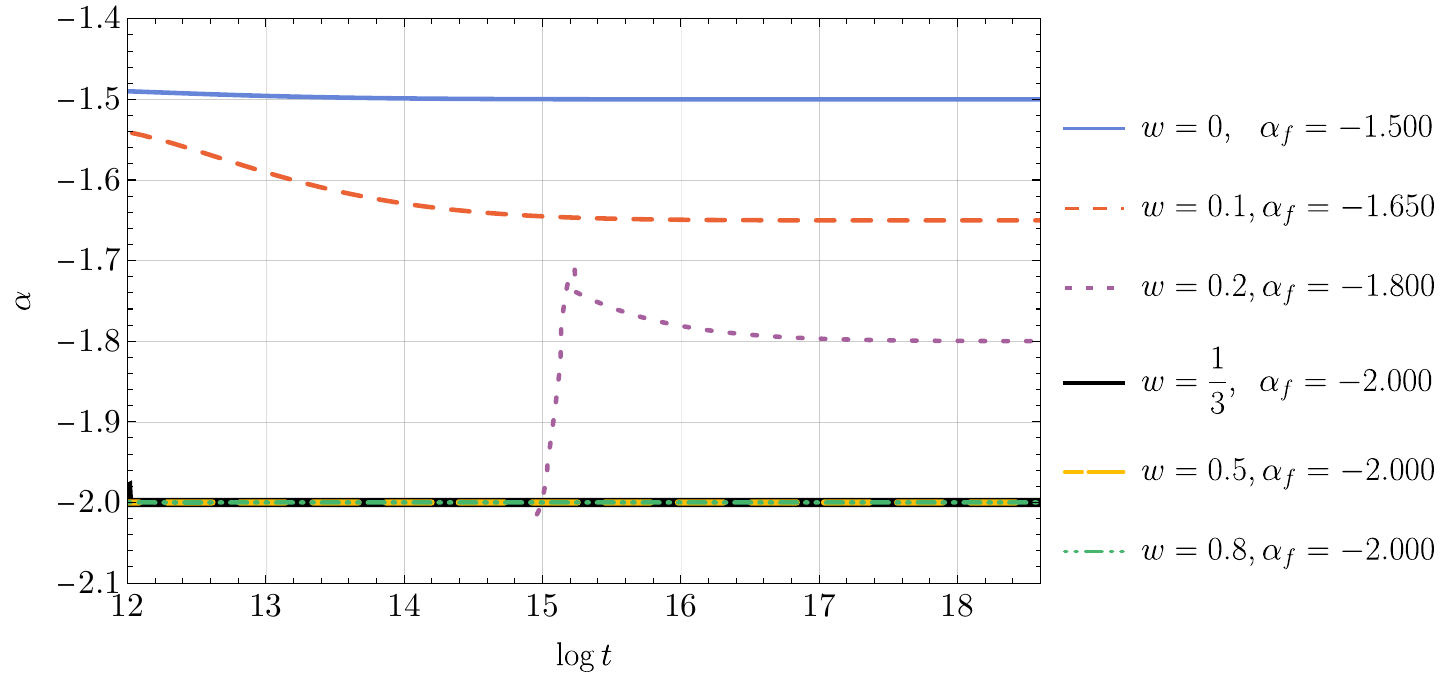}
\caption{$\sup_{\mathbb{H}^3}|\phi|$ decays as $t^{\alpha}$ for $w \ge 0$. Initial data \eref{inidata}, $dt=0.01$, $dr=0.05$, $r \in [0, 100]$. The value $\alpha_f$ represents the final value of each plotted line.}
\label{fig:hyp_phi_globaldecay}
\end{figure}

We now consider the background metric
\begin{equation*}
g = -dt^2 +a(t)^2 \left(dr^2 + \sinh(r)^2\left ( d\theta^2 + \sin^2(\theta) d\varphi^2 \right) \right)
\end{equation*}
 and evolve the scalar field using the conformal time,  that is, by using \eref{wave_hyp_conformal} together with the scale factor $a(\tau)$ given by \eref{scalefactor_hyp}.
The following decay rates are readily obtained (see figure \ref{fig:hyp_phi_globaldecay}):
\renewcommand*{\arraystretch}{1.2}
\begin{equation} \label{hyp_decay_phi}
 \sup_{x \in \mathbb{H}^3}|\phi|(t, x) \sim \left\{
\begin{array}{ll}
t^{-\frac32(w+1)}, &\textrm{if } 0 \le w \le \frac13, \\
t^{-2}, &\textrm{if } w \ge \frac13.
\end{array}
\right.
\end{equation}

\begin{figure}[h!]
\centering
\includegraphics[width=0.8\textwidth]{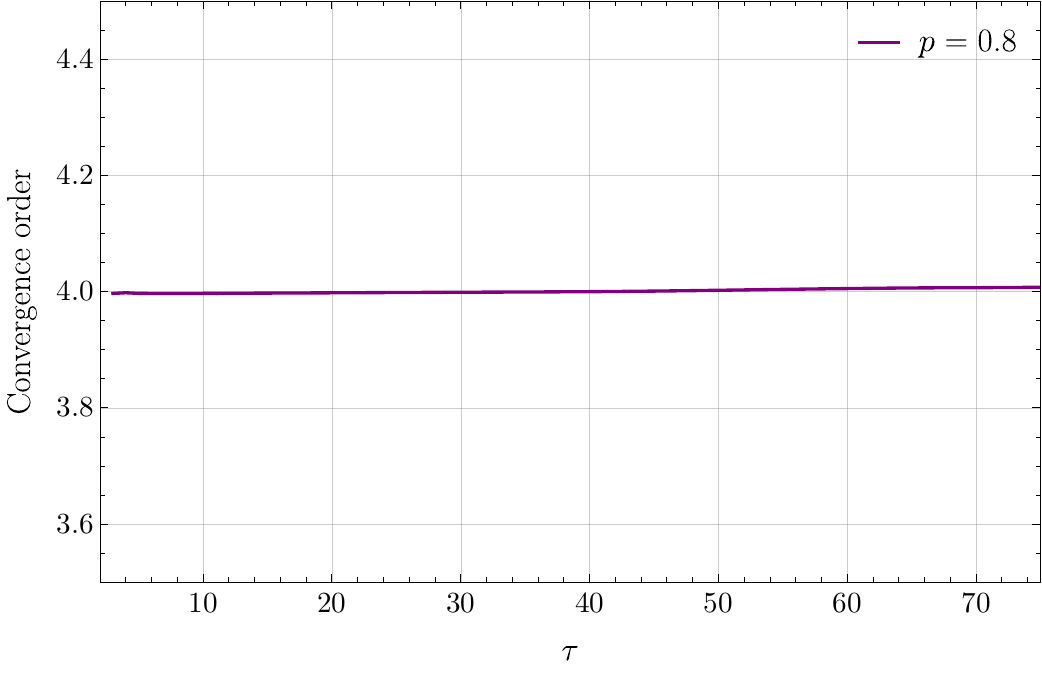}
\caption{$L^2$ norm convergence plot for the solution $\phi$ in a flat FLRW universe with $p=0.8$.}
\label{fig:flat_phi_convergenceplot0.8}
\end{figure}

When $w=0$ (dust solution), in particular,  this confirms that $\phi$ decays as $t^{-\frac32}$ for large times, as proved in \cite{NatarioRossetti}. In view of the near perfect 4th order convergence of our solutions (see figure \ref{fig:flat_phi_convergenceplot0.8}), such a numerical counterexample also stands as a strong evidence against the decay rate $t^{-2}$ stated in \cite{AbbasiCraig} for this specific FLRW universe. More generally, we could  observe that the solutions decay at a rate slower than $t^{-\frac32(w+1)}$ for $-\frac13 < w < 0$ and that there is no decay to zero for $-1 < w < -\frac13$. 

The results of figure \ref{fig:hyp_phi_globaldecay} can be interpreted as follows: at the initial time, the maximum of the solution is given by the peak of the initial Gaussian curve. In the case $w=0$, this peak approaches zero but stays the maximum of the solution for the entire dynamics. For $w > 0$, however, there is a competition between the tail at the origin and the contribution given by outgoing waves. For $0 < w < \frac13$, the former element (the tail) will be seen to give the slowest contribution to the decay after a finite time $T$, but the closer $w$ gets to $\frac13$, the larger is $T$.  For $w \ge \frac13$, the dominant (i.e. slowest) contribution is given by the outgoing waves, and therefore the tails have a secondary role in the asymptotics. This behaviour is analogous to the one occurring in the flat case due to the interplay between redshift and dispersion. 

Furthermore:
\begin{equation} \label{hyp_decay_pi}
 \sup_{x \in \mathbb{H}^3}|\partial_t \phi|(t, x) \sim \left\{
\begin{array}{ll}
t^{-\frac32(w+1)-1}, &\textrm{if } 0 \le w \le \frac13, \\
t^{-3}, &\textrm{if } w \ge \frac13.
\end{array}
\right.
\end{equation}
In the current case, the decay of $|\partial_t \phi|$ as a function of $w$ is determined by the same intervals present in \eref{hyp_decay_phi}. Indeed, although $\partial_{\tau} \phi$ solves a wave equation which is different from the one satisfied by $\phi$, both equations have time-dependent coefficients having the same asymptotics. This is due to the fact that  $a(\tau)$ has an exponential profile for large times, which is a shared property among the spacetimes having $w \ge 0$.

As discussed in the previous section, the Huygens principle is an unstable phenomenon which is generally expected not to hold. It does apply in the radiation-filled universe ($w=\frac13$), also in the hyperbolic case. After excluding the latter spacetime and other possible isolated cases for which the principle holds, we found numerically that for every $w  \ge 0 $ we have, for every $\delta \in (0, 1)$:
\begin{equation*}
\sup_{ |x|<\delta \tau } |\phi| \sim 
t^{-\frac32(w+1)}
\end{equation*}
and
\begin{equation*}
\sup_{ |x|<\delta \tau } |\partial_t \phi| \sim 
t^{-\frac32(w+1)-1}.
\end{equation*}
Outside such a lightcone, the asymptotic profiles for $\phi$ and $\partial_t \phi$ are as in \eref{hyp_decay_phi} and \eref{hyp_decay_pi}, respectively. 


It is interesting to notice that, when comparing solutions propagating in spacetimes of different spatial curvature, but characterized by the same speed of sound, the hyperbolic case is associated to faster decay rates. This is due to the  stronger dispersive effects ensured by the hyperbolic geometry of space.

Finally, and as evidence that our results are reliable, we point out that we obtained nearly perfect 4th order convergence in our numerical experiments, like in the example configuration shown in figure~\ref{fig:flat_phi_convergenceplot0.8}.

\section{Conclusions}

The numerical study of the wave equation in expanding FLRW spacetimes presented here yielded new results while, at same time, described the precise mechanism behind the known decay rates. In the flat case, the asymptotics that we obtained for the wave solutions and their time derivative provided evidence for the optimality of the acknowledged estimates  \cite{NatarioRossetti}. Moreover, we broke down the different decay contributions to find that tails, namely backscattered waves detected inside the lightcone, represent the dominant contribution to the global asymptotics when $\frac23 < p < 1$. We hope that this behaviour might be observed and exploited analytically through local energy decay bounds, in the style of the Morawetz estimate in Euclidean space \cite{Morawetz}. This might be relevant to develop a full mathematical treatment that can be robust under non-linear perturbations. Indeed, although upper bounds for wave solutions have been determined rigorously, the proofs for the region $0 < p < 1$ (corresponding to the presence of a null \scrip) rely on Fourier analysis and are therefore not easily adaptable to non-linear problems.  

In the spatially-hyperbolic spaces, the numerical results filled an even larger gap. Indeed, we obtained precise decay rates for the region $w \ge 0$, thus complementing the results known for the dust-filled ($w=0$) and the radiation-filled ($w=\frac13$) universes only.  Except for these two cases, a comprehensive mathematical treatment of this problem is missing, even with Fourier methods, also due to technical difficulties resulting from the curvature of spatial sections. Our numerical results on the global decay and on behaviour of tails (that we showed to be dominant for $0 \le w < \frac13$) can therefore provide useful insights to face the problem in full generality.

In the current setup, the background where the scalar field propagates is sliced into Cauchy  hypersurfaces cut at a given value of the radial coordinate. A disadvantage of this approach is that the boundary conditions imposed at the outer boundary introduce spurious reflections travelling inwards and thus limit in time and radius the reliability of the numerical data evolved. Hyperboloidal slices \cite{Zenginoglu:2007jw} are spacelike hypersurfaces that reach future null infinity \scrip, which in the cases considered here is an ingoing null surface that will not allow any physical information to enter. By radially compactifying a hyperboloidal slice, the outer boundary of the domain can be set at \scrip $ $ without the need of any boundary conditions (see e.g. \cite{Vano-Vinuales:2014koa}). Our plan is to apply the hyperboloidal approach to the cosmological scenarios considered here. 

\ack

The authors thank João Costa, David Hilditch and José Natário for useful discussions and valuable comments on the manuscript.
FR was supported by FCT/Portugal through the PhD scholarship UI/BD/152068/2021, and partially supported by FCT/Portugal through CAMGSD, IST-ID,
projects UIDB/04459/2020 and UIDP/04459/2020.
AV thanks FCT for financial support through Project~No.~UIDB/00099/2020. 
\section*{References}
\bibliographystyle{unsrt}
\addcontentsline{toc}{chapter}{References}
\bibliography{../refs}

\end{document}